# ANALYSIS OF SYSTEM CAPACITY AND SPECTRAL EFFICIENCY OF FIXED-GRID NETWORK

Adarsha M[1], S. Malathi[1], Santosh Kumar[2]

[1]Department of Electronics and Communication Engineering, M. S. Ramaiah University of Applied Sciences, Bangalore-560054, India
[2]Shandong Key Laboratory of Optical Communication Science and Technology, School of Physics Science and Information Technology, Liaocheng University, Liaocheng - 252059, China

## ABSTRACT

*In this article, the performance of a fixed grid network is examined for various modulation formats to estimate the system's capacity and spectral efficiency. The optical In-phase Quadrature Modulator (IQM) structure is used to build a fixed grid network modulation, and the homodyne detection approach is used for the receiver. Data multiplexing is accomplished using the Polarization Division Multiplexed (PDM) technology. 100 Gbps, 150 Gbps, and 200 Gbps data rates are transmitted under these circumstances utilizing various modulation formats. Various pre-processing and signal recovery steps are explained by using modern digital signal processing systems. The achieved spectrum efficiencies for PM-QPSK, PM-8 QAM, and PM-16 QAM, respectively, were 2, 3, and 4 (bits/s)/Hz. Different modulation like PM-QPSK, PM-8-QAM, and PM-16-QAM each has system capacities of 8-9, 12-13.5, and 16-18 Tbps and it reaches transmission distances of 3000, 1300, and 700 kilometers with acceptable Bit Error Rate (BER$\leq 2 \times 10^{-3}$) respectively. Peak optical power for received signal detection and full width at half maximum is noted for the different modulations under a fixed grind network.*

## 1. INTRODUCTION

An Optical Network is a communication network used for the exchange of information through an optical fiber cable between one ends to another. It is one of the quickest networks used for data communication[1]. Optical networks offer increased capacity and reduced costs for new applications such as the Internet, video and multimedia interaction, and advanced digital services [2]. Global demand for high data rates is rising, so researchers are looking for different ways to supply gigabit capacity [3]. There are various classes of optical communication networks. For instance, multiple wavelengths per optical fiber network architecture are used in central, metropolitan, or wide-area applications to connect thousands of users with a wide range of transmission speeds and capacities. Sending multiple wavelengths through a fiber 1300 to 1600 nm range at the same time is a powerful feature of an optical communication link [4][5]. The technology of combining multiple wavelengths onto a single fiber is known as wavelength division multiplexing (WDM)[6]. The use of the WDM principle in conjunction with optical amplifiers results in communication links that allow users from all over the world to communicate quickly [7]. The outdated 10 Gbps transport optical networks were updated to 40–100 Gbps networks to accommodate the varying bandwidth requirements of a variety of services [8]. Within the same fiber, simultaneous transmission of 10/40/100 Gbps on various wavelengths is possible.

The bulk discount on high-bit-rate transponders could lower the overall cost of transmission[9]. In fixed-grid networks, a particular kind of transceiver is selected, and it serves a single demand. It fixes the data rate, range, and spectrum used. Fixed networks use 50 GHz channel spacing because they are based on the fixed grid as specified by the ITU-T[10]. Commercial optical fiber communication systems' transmission capacities have been growing at a rate of 140% annually. The trend is likely to persist due to the anticipated demand sparked by the launch of new data communications and high-definition video services[11]. The introduction of commercial transport networks typically occurs 5 to 10 years after the relevant study, and their transmission capacities have also continued to rise[12]. If the current growth rate is maintained, the commercial capacity will reach its limit within the next ten years. At present Time Division Multiplexing (TDM), and WDM technologies are introduced to meet the demand for more capacity and are used to multiplexes several channels. To handle higher modulation impairment Digital Coherent Transmission (DCT) technologies are used to provide additional spectral efficiency[13]. WDM and TWDM [14] will be the top two technology choices for back-haul and front-haul access network design and deployment.

## 1.1. Motivation

By 2022, it's projected that traffic between machines and connected devices would increase globally at a compound annual growth rate of 47%[15][16]. By 2023, Cisco predicts that 50 billion people using fixed-line, mobile, and machine-to-machine internet will need between 44 and 110 Mbps of bandwidth per user to access modern applications[17] [18]. All this prediction shows that there is high demand for data rates so which drives to do the investigation of the C-band capacity for upcoming demand.

## 1.2. Problem Statement

One can go for the very expensive cable like EDFA for better Optical Signal Noise Ratio(OSNR) as well as spectral efficiency. Replacement of under-laying cable is very cost-consuming as well as labor-intensive. Because of that traditional network architecture setup is kept as it is, to improve the system capacity. Additionally, the following major research gaps are addressed in this paper:

- What opportunities are available to boost C-band spectral efficiency and system capacity?
- It investigates the peak optical power requirements for the different modulation schemes in fixed grid networks.
- Which information coding gives the best BER rate for different modulations?

## 1.3. Paper Contribution

Considering the prediction Fixed grid optical network is constructed and simulated in this article to analyze spectrum efficiency and system capacity.In the transmitter section, differential coding and gray coding are introduced. As a result, BER effectiveness is evaluated and compared. Additionally, various modulation schemes are transmitted and its OSNR versus BER are observed.

The rest of this article is broken down into the sections that follow. Evolution of Optical Communication Network (OCN) and related work in section 2. Introduces details of the fixed grid in section 3. The design of the fixed-grid network is presented in section 4. The simulation of fixed-grid networks is described in section 5. In section 6, the system capacity and spectral efficiency of various simulation results obtained from fixed-grid networks are presented and discussed. Finally, the conclusions of the study are presented in section 7.

## 2. EVOLUTION OF OPTICAL COMMUNICATION NETWORK

When optical fibers were first utilized for optical communications in 1977, the slope changed. The bit-rate–distance product $BL$, where $B$ is the bit rate and $L$ is the repeater spacing, the distance after which an optical signal must be regenerated to maintain its fidelity, is a commonly used figure of merit for communication systems[19]. Table 1 indicate how technological advancements have raised from 1977 to 2015. The first generation of Optical Communication (OC) systems used GaAs semiconductor lasers with wavelengths of about 850 nm inside their optical transmitters. Before reaching an optical receiver, the optical bit-stream was transferred over graded-index multimode fibers and transformed to the electric domain using a silicon photodetector.System designers were motivated by the larger repeater spacing compared to the coaxial system's 1 km spacing since it reduced the installation and maintenance expenses associated with each repeater. One can observe that attenuation gradually decreases from generation to generation which allows signals to travel for long distances without using the repeater.

Table 1 Evolution of Optical Communication Network

| Megabits per second | | | |
|---|---|---|---|
| **Details** | **1ˢᵗ G** | **2ⁿᵈ G** | **3ʳᵈ G** |
| Optical Fiber | Multi-Mode Fiber | Single Mode Fiber | Single Mode Fiber |
| Link /Topology | P-T-P link | Ring, P-T-P | Ring, P-T-P |
| Multiplexing | Bit-wise multiplexing (TDM) | Byte-wise multiplexing (TDM) | Bit-wise multiplexing (TDM) |
| Data rate | 1-45 Mbps | 50 M bps – 2 G bps | 50 Mbps – 10 Gbps |
| Reach | 10Km | 50Km | 50Km |
| Attenuation | 3 dB/km | 1 dB/km | 0.2 dB/km |
| Year | 1977-1980 | 1981-1987 | 1988-1994 |
| **Gigabit per second** | | | |
| **Details** | **4ᵗʰ G** | **5ᵗʰ G** | **6ᵗʰ G** |
| Optical Fiber | Single Mode Fiber | Single Mode Fiber | Single Mode Fiber |
| Link /Topology | Ring, Mesh | Ring, Mesh | Ring, Mesh |
| Multiplexing | WDM | WDM /PDM | WDM /PDM |
| Data rate | 2.5 G bps – 40 G bps | 10 G bps – 100 G bps | 10 G bps – 200 G bps |
| Reach | 50Km | 50Km | 50Km |
| Attenuation | 0.2 dB/km | 0.2 dB/km | 0.2 dB/km |
| Year | 1995-2001 | 2001-2010 | 2010-2015 |

P-T-P: Point To Point link          G: Generation

It's worth noticing that from first to third generation TDM is used. By 1990, system designers had shifted their attention to system capacity, fiber loss, and fiber dispersion. The fiber loss is overcome by periodic optical amplification similarly fiber dispersion is controlled by periodic dispersion compensation. Finally, system capacity is handled by using WDM and PDM. In 2011, 64-T bit/s transmission over 320 km of single-mode fiber was achieved using 640 WDM channels spanning both the C and L bands with 12.5-G Hz channel spacing. Each channel has two polarization-multiplexed 107-G bit/s signals that were coded using the Quadrature Amplitude Modulation (QAM) scheme [20]. The greatest capacity Discrete Multi-tone Modulation (DMT)signal transmission across 2.4-km SMF with the BER under SD-FEC constraint of 2.4 $\times 10^{-2}$ was accomplished using a four-channel WDM 256.51Gbps 16-QAM-DMT short-reach optical-amplifier-free interconnection [21].

## 2.1. Related Works

In the present generation (2015 – 2022) to overcome transmission losses, digital coherent receivers are used[22]. The digital coherent receiver relies on robust digital carrier-phase estimation, one can use several spectrally effective modulation patterns[23]. Furthermore, the phase information is kept after detection, we can use Digital Signal Processing (DSP) [24] to equalize linear transmission impairments such as Group-Velocity Dispersion (GVD)[25] and Polarization-Mode Dispersion (PMD)[26] of transmission fibers. Recently, 100-Gbit/s transmission systems using Quadrature Phase-Shift Keying (QPSK)[27] modulation, PDM, and phase-diversity homodyne detection aided by high-speed DSP have been developed and used in commercial networks[28]. It is possible to compensate for chromatic dispersion by periodically deploying special optical fibers known as Dispersion Compensating Fibers (DCF)[29], allowing amplifiers and DCFs to handle multiple wavelengths simultaneously.

Table 2 Different Challenges in OCN and Its Solutions

| Challenges | Solutions | References |
|---|---|---|
| Speed, Quality | Coherent detection polarization multiplexing, digital processing, and multilevel modulations | [31][29] |
| Performance and cost | Providing the most prominent factors of the Radio Over Fiber (ROF) architecture reduces the system installation costs. | [32] |
| Fiber Nonlinearity | Estimating nonlinear noise power and OSNR induced via fiber nonlinearity by Long Short-Term Memory (LTSM) network. | [33] |
| Reducing the communication degradation | Adaptive Field/ Digital Signal Processing | [34][26] |
| Handling nonlinearity in Single-channel optical communication | Splitting the nonlinearity compensationis always advantageouswhen there are two or more spans. | [35] |
| To Achieve ultrahigh-capacity fiber communications | Effective coherent multi-wavelength sources are used for the new generation of coherent fiber communication networks. | [36] |
| To overcome data center network traffic | Photonic Integrated Circuits (PIC), improved fiber optic communication infrastructure, using the full spectrum range of fiber optic technologies, and signal modulation to resolve losses are used. | [37][38] |
| Canceling Kerr-induced transformations to increase the capacity | Fiber information capability can be substantially improved concerning previous estimates. | [39] |
| Nonlinearity in coherent optical communication | Proposed and demonstrated a basic nonlinear equalizer based on the Functional-Link of Neural Networks (FLNN). | [40] |

Modern WDM optical networks rely on spans to connect nodes. These are 70-100 km fiber lengths with amplifiers Erbium-Doped Fiber Amplifiers (EDFAs)[30] and DCFs. The total transmission range could be several thousand kilometers in this manner (without O-E-O conversion). The demand for bandwidth is increasing due to the explosive growth of Internet services such as video conferencing, Net-Fix, cloud computing, and mobile access with video clients. This requires an expansion of the transmission capacity of optical fibers and the development of next-generation high-speed optical networks. The various challenges faced by OCN are listed in Table 2 along with possible solutions. Where it addresses the network speed, quality, performance, and cost. To reduce communication degradation DSP is utilized in the receiver end. To enhance the network performance PIC is incorporated which makes an easy way to do the signal modulation and to resolve losses in the transmitter side.

Unlike in the early days of DWDM systems, when an optical fiber's bandwidth was thought to be unlimited, the optical spectrum [41] will be a valuable commodity now in data centers, and the industry is now looking for ways to increase overall spectrum efficiency. Considering these facts we tried to explore increasing the spectral and system capacity. One way to overcome this issue is to have error-free communication and also use WDM. Therefore, we have implemented the channel coding in the transmitter part to reduce bit error rates. Both differential and gray coding are adopted for information encoding. Differential coding involves encoding data by representing the difference between consecutive values or samples rather than encoding each value independently. This can be particularly useful when the variations between successive data points are smaller compared to the absolute values themselves. The primary goal of differential encoding is to reduce redundancy in data representation, which can lead to more efficient storage and transmission. It is especially effective when dealing with data that exhibits smooth or gradual changes over time. The primary advantage of Gray code is that it reduces the likelihood of errors when transitioning from one value to the next. In traditional binary representation, transitioning from one value to another can result in multiple bits changing simultaneously, which can lead to errors due to timing or noise. In Gray code, only one bit changes at a time, which reduces the chances of errors during transitions. Such a coding system is adopted for different modulations. Grey and differential codes systems OSNR requirements are observed. By keeping transmission distances, data rates, and fiber launch optical powers are used for the modulation formats are taken into consideration. Overall C-band capacity is investigated using different modulation techniques and also analyzed its spectral efficiency including each modulation max reach.

## 3. DETAILS OF THE FIXED GRID NETWORK

The Conventional band (C band) has the lowest losses across the spectrum, so this band is used to transmit data over an extremely long distance. The C band includes wavelengths between approximately 1525 and 1565 nm. Wavelength allocation and standardization were set by ITU-T. A technique called Dense Wavelength Division Multiplexing (DWDM) is used. It makes it possible to transmit numerous optical signal carriers at various wavelengths through a single optical fiber. DWDM central frequencies are specified in the ITU-T G.694.1 guideline [42].

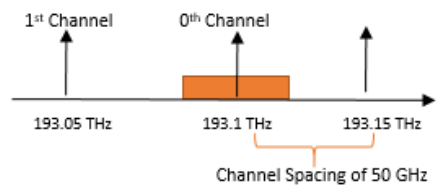

Figure 1. Fixed grid channel spacing

DWDM wavelengths were first placed in a grid with an optical frequency spacing of exactly 100 GHz which is approximately 0.8 nm wavelength. Over the past ten years, numerous significant new developments have increased the capacity to keep up with the steadily growing traffic.

For instance, core networks can transmit about 80 channels by compressing the channels and spacing that 50 GHz apart. To build a fixed grid network 50GHz channel spacing is used to transmit data rate. 193.10 THz will serve as the reference frequency for this design. A single carrier network is made up of a fixed grid (50 GHz) that transmits either a single line rate or a mixed line rate. Figure 1 illustrates the fixed grid network channel spacing. Capacity scaling is possible in a communication system by exploring modulation so Quadrature Phase Shift Keying (QPSK) and Quadrature Amplitude Modulation (QAM) are explored. To increase the bit rate per second Dual-Polarizations (DP) technique is adopted[29].

## 4. DESIGN OF FIXED GRID NETWORK

The fixed grid network is shown in Figure 2 it consists of a transceiver, Optical Cross-Connect (OXC), and optical fiber. Different clients are connected by using mesh topology in this example. It's also possible to connect different clients with different topologies. An enhanced form of an optical network called a fixed grid optical network is made to deliver dependable, fast communication between numerous locations. They offer a practical method for developing an optical network without the use of pricey and complicated routing methods. Based on a mesh network architecture, fixed grid optical networks connect each node to several other nodes in a preset grid pattern. Each node in the network is directly connected to several other nodes, enabling effective communication between all of the nodes. High scalability, low latency, and dependable data transmission are benefits of employing a fixed-grid optical network. They are therefore perfect for services like streaming media, video conferencing, and long-distance communication.

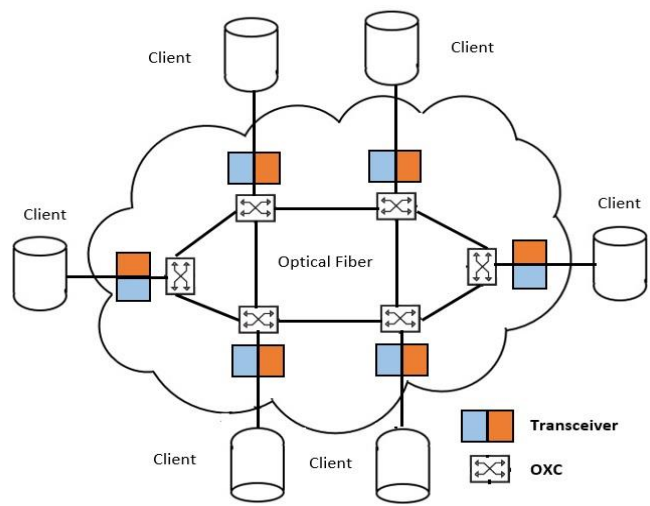

Figure 2. The architecture of fixed grid network

### 4.1. Fixed Gird Network Transceiver

A transceiver is a transmitter and receiver combined in a single unit. Figure 3 gives the standard design schematic of the fixed grid transmitter. Which exhibits the complete design configuration of the DP-QPSK modulation formats. The transmitter unit divides the input bit sequence in half evenly using a serial-to-parallel converter. Both even and odd parts are present in each sequence. For the QPSK, each bit sequence is transformed from binary signals into M-ary symbol sequences using Phase Shift Keying (PSK). Phase ambiguity is eliminated with QPSK/QAM modulation methods utilizing differential encoding[43]. The multilayer pulse is produced by the M-ary pulse generator. M-Ary pulse generator out will drive the IQ modulators I and II. An IQ modulator comprises two-phase modulators, a phase shifter, and two couplers cross couplers. Mach–Zehnder Modulator (MZM) is used to design the IQ modulator which works under the push-pull configuration.

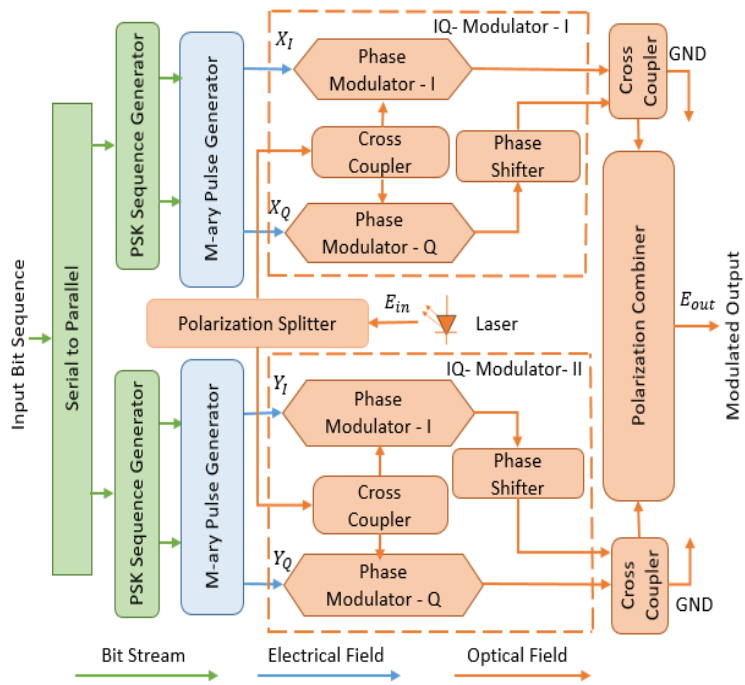

Figure 3. Fixed grid transmitter

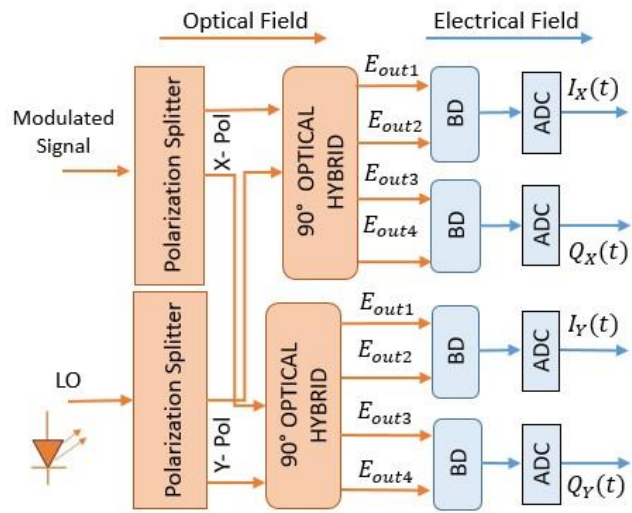

Figure 4. Fixed grid receiver

Continuous light is used as the input, and a Polarization Splitter (PS) divides it into two orthogonally polarized, equally powerful beams of light. An IQ modulator is used to modulate the two orthogonal polarization lights. Each phase modulator handles data streams coming from the M-ary pulse generator. At the end of the IQ-modulator, the modulated signal is going to be combined with the help of a polarization combiner so the modulated output is ready for communication. The DP-QPSK/DP-QAM optical signal is demodulated using a coherent detection technique which is shown in Figure 4. The incoming data stream is divided into two by the PS. Two 90° optical hybrids are used to combine the signal, which has both X and Y polarization, with the local oscillator (LO)[1]. Information about intensity is generated from phase difference information by combining the optical carrier-containing data with the LO signal. The Balance Detector's (BD) light is transformed into analog signals. A high-speed Analog to Digital Conversion (ADC) sampling transforms such signals into digital signals. After being precisely sampled, the signal can be recovered.

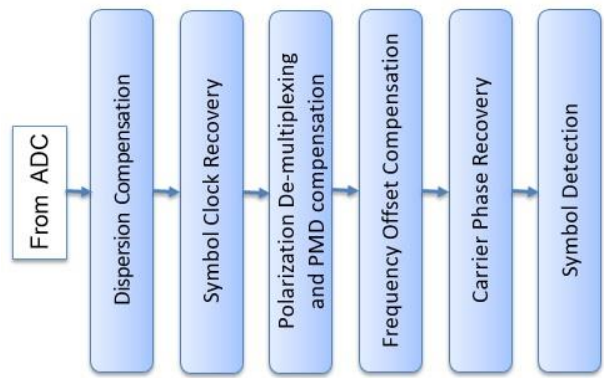

Figure 5. DSP unit in the Receiver section

To overcome the loss that occurred from the transmitter to the reception section, a Digital Signal Processing (DSP) unit is used which is shown in Figure 5. The leading cause of signal transmission loss in optical fiber is Chromatic Dispersion (CD) [44][26]. The frequency domain is used to do CD compensation since it requires less computation when the compensation value is higher. As a result, data is first translated into the frequency domain, followed by multiplication by the inverse transfer function of the dispersion function, and finally, turn backs to the time domain. After CD compensation it enters the symbol clock recovery to overcome clock misalignment. Due to the independence of the transmitter output data clock and the A to D sampling clock, the clock's frequency and phase can differ. The symbol clock recovery algorithm determines the symbol clock frequency, decides on the best sampling point, and then resamples the data appropriately.

---

[1] Local Oscillator: This is nothing but the laser, having similar properties as the source laser especially line width almost equal to the source laser.

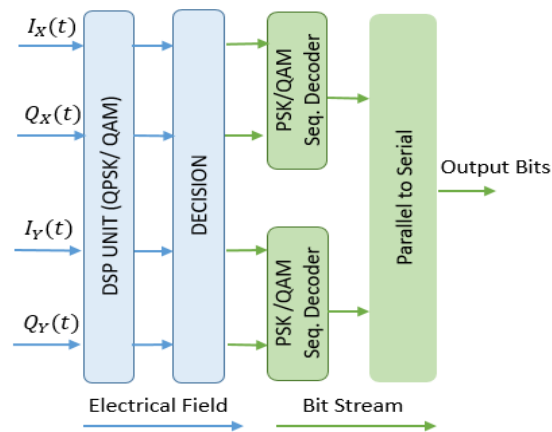

Figure 6. The primary instance of demodulation

It follows the double sampling rules which mean 2 samples/symbol. Thus, by doing this, it is possible to maintain synchronization between the sample clock and the receiver's launching symbol clock. Following the acquisition of the appropriate symbol clock for each polarization. There are two signals (one on each polarization) but due to the polarization's rotated state, the data is mixed. Polarization Mode Dispersion (PMD) can be resolved by using the inverse Jones matrix[2]. It will separate the signal's two orthogonal polarizations and compensate for signal loss. The Constant Modulus Algorithm (CMA) is used to compensate for the PMD. As a result, two polarization signals are apart but rotating as a result of carrier frequency offset and phase noise. The same theory underlies carrier phase recovery and Frequency Offset Compensation (FOC). Estimating the constellation diagram's rate of rotation is intended to be followed by the offset's removal. It is performed with the help of the Viterbi–Viterbi algorithm[44].

Finally, by applying the proper thresholds to the received constellation slices, one can identify the symbols (and bits) transmitted through the fiber. Depending upon the coding technique applied in the transmitter section an appropriate reverse de-coding needs to be used to extract the bit stream, which is shown in Figure 6. Reversing the differential encoding process is known as differential decoding. Taking into account the variations between succeeding values, it entails recovering the original data from the encoded or differentially encoded values.In the binary numbering scheme known as gray code, which is also referred to as reflected binary code or unit distance code, two successive values are separated by only one bit. Recovery of the original binary values from their Gray code is known as gray decoding. M-ary pulse generator involves both differential and gray coding techniques which is encoding the PSK sequence signal. Similarly, PSK/QAM decoder will extract the original signal from encoded data. Both differential and gray coding and decoding are introduced in this work and analyze the BER and OSNR.

## 4.2. Optical Cross-Connect (Oxc)

In the Older days, electronic signals are mostly responsible for all networking equipment's operation. That first optical signal was transformed into electrical, then it was amplified, regenerated, or switched before being converted back into optical signals [45]. For all varieties of optical networks, OXC is the most appealing key component. It switches at a very fast rate and with good reliability[46]. In networks, wavelength routing is provided via OXC, which is used to connect any two topologies. There are two different types of OXC switches one is the digital switch which is opaque or hybrid the second one is transparent OXC[47].

---

[2]Jones matrix: It gives the details of different polarization states.

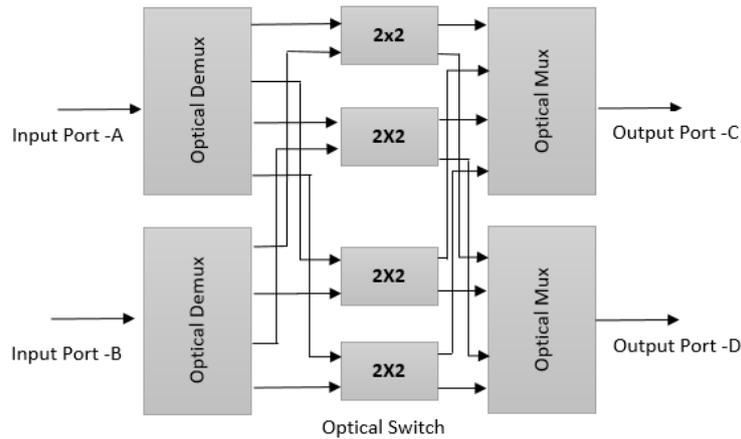

Figure 7 OXC implemented using 2X2 Optical switch

With the use of electronic cross-connection technology, optical data streams are first changed into electronic data in the digital OXC switch, which is then transformed back into optical data streams. OXC operates specifically in the photonic field[48]. Figure 7 illustrates the creation of an OXC in optical fiber communications using de-multiplexed and multiplexed WDM channels.

## 5. SIMULATION OF FIXED GRID NETWORK

Based on the OptiSystem V.18 simulation platform, the reference model of the fixed grid network is constructed as depicted in Fig. 2. The hypothetical center frequencies that are permitted are represented by a grid of frequencies. For DWDM systems, six channels with equal channel separations of 50 GHz are utilized. Fixed grid nominal central frequencies are supplied by 193.1 + n × 0.05, where n is a positive or negative integer including 0 [5]. The central frequency of each channel is assigned which is mentioned in Table 3. The addition of optoelectronic components that operate at higher data rates while staying within the 50 GHz grid has increased channel capacity.

Table 3 Central frequency allocation for each channel

| Channel Number | Nominal central frequencies (THz) for spacing of each channel |
|---|---|
| 1 | 193.1 |
| 2 | 193.15 |
| 3 | 193.2 |
| 4 | 193.25 |
| 5 | 193.3 |
| 6 | 193.35 |

In this design, the fiber loss is assumed to be 0.2 dB/km. The dispersion is 0.2 ps/nm.km. When the data rate is increased from 2.5 to 100 Gbps per wavelength while channel spacing remains constant. To generate 100Gbps, Dual-Polarization-Quadrature Phase Shift Keying (DP-QPSK) modulation scheme with 25 G baud rate and 2 Bits/Symbol is used.

Table 4 Fixed grid network Simulation parameters

| Parameters Values | PM-QPSK |
|---|---|
| bit rate | 100 Gbps |
| baud rate | 25 G-baud |
| Fibre attenuation coefficient | 0.2 dB/km |
| Fibre dispersion coefficient | 17 ps/nm/km |
| Fibre differential group delay | 0.2 ps/km |
| laser power | 14 dBm |
| laser central wavelength | 1550 nm |
| laser linewidth | 0.1MHz |
| laser initial phase | 0° |
| fiber launch power | 0 dBm |
| EDFA gain | 8 dBm |
| EDFA noise BW | 4 THz |
| EDFA noise figure | 6 dB |
| photodetector responsivity | 1 A/W |
| photodetector dark current | 10 nA |

The fixed grid optical network simulation parameter is expressed in Table.4 which involves the channel impairments, bit rate, baud rate, receiver sensitivity, and EDFA gain and noise figure. A similar setup is carried out for the PM-M-QAM modulation scheme. Where 150 Gbps data rate is generated by using PM-8-QAM modulation scheme with 25 G baud rate and 6 bits/symbol is used. With the 50GHz channel spacing maximum, 200 Gbps data rate is transmitted by using PM16-QAM scheme with 25 G baud rate with 8 bits/symbol used.

# 6. RESULT ANALYSIS

The QPSK modulation scheme is taken into consideration for a better understanding of and visualization of the constellation diagram in the different stages of the DSP receiver section. After dispersion compensation and non-linear compensation by acquiring the proper symbol clock for each polarization, the X and Y-polarization constellation diagrams are shown in Figure 8(a). We receive two signals, one for each polarization, but due to the rotated state of the polarization, they have mixed data. These problems are addressed by PMD compensation. The constellation diagrams after the Polarization De-multiplexing and PMD compensation which is expressed in Figure. 8(b). Where two signals are separated but they are rotating due to carrier frequency offset and phase noise. It can be overcome by using FOC and Carrier Phase Estimation (CPE) Figure8(c) shows the respective constellation diagrams. Finally, by applying the proper thresholds to the received constellation slices, we can identify the symbols (and bits) that were transmitted via the fiber.

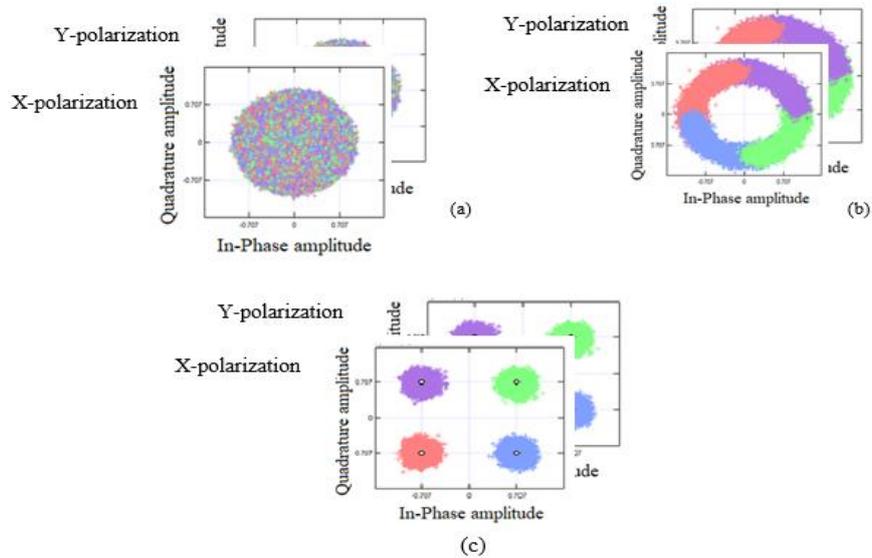

Figure 8. Constellation diagrams at various DSP component levels. (a) After dispersion compensation and right symbol clock for each polarization (b) After Polarization De-multiplexing and PMD compensation (c) After FOC and Carrier Phase Recovery

Figure 9 shows the transmitter output for three polarization multiplexed gray-coded optical modulation formats that were evaluated. For PM- QPSK, PM- 8-QAM, and PM- 16-QAM the reported peak optical powers are -5.20962, -10.6693, and -14.63234 dBm respectively. The Full Width at Half Maximum (FWHM) is also observed to be 20.6, 14.6, and 10.7. An IQ Modulator is used in optical communication for transmitting data in the form of light pulses. It modulates light waves by combining in-phase and quadrature components of the data signal, resulting in a composite signal. This composite signal is then transmitted through an optical fiber. PM-16 QAM is an IQ Modulator that combines sixteen in-phase and quadrature components to generate sixteen light pulses. It has a very wide bandwidth and is capable of very high data rates. It is used in high-speed optical communication systems. By altering the M-Ary pulse generator, the IQmodulator can function as PM-QPSK and PM-QAM.

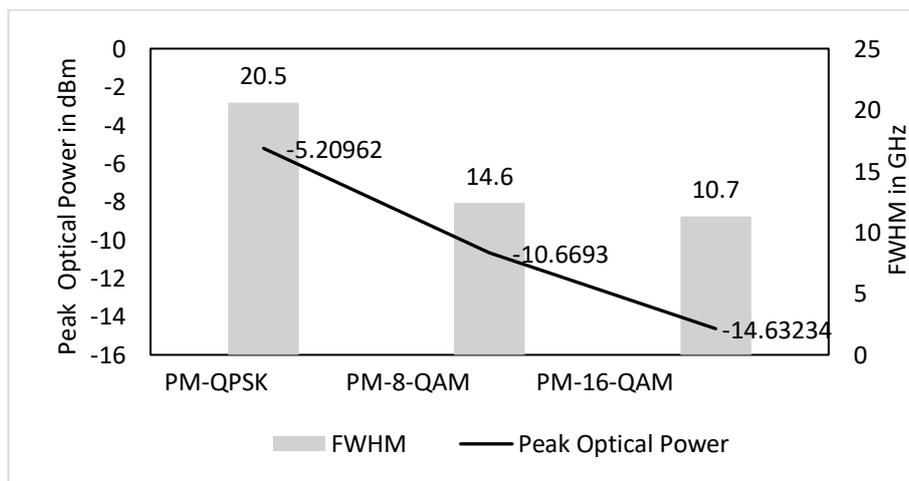

Figure 9. Different modulation schemes its peak Optical power and FWHM.

The term OSNR describes the proportion of a transmission link's signal optical power to its noise optical power which is expressed in Eq.1[13]. The level of optical noise interference on optical signals inside a valid BW is measured using OSNR. The OSNR requirement is inversely related to the Euclidean distances between the constellation points for different modulation schemes.

$$OSNR(dB) = 10 \log \left( \frac{P_{signal}(mW)}{P_{noise}(mW)} \right) \qquad \text{(Eq.1)}$$

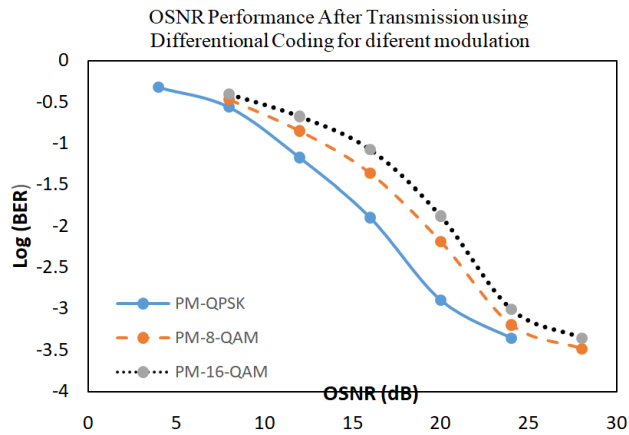

Figure 10 OSNR Performance of Transmitted Signal using Differential Coding

Hence, with a given BER ($\leq 2 \times 10^{-3}$), the OSNR tolerance decreases as the modulation format increases which is depicted in Figures 11 and 12. For systems using various modulations with differential and gray-coding sequences, BER is determined concerning the user-defined OSNRs.

For PM-QPSK, PM-8-QAM, and PM-16-QAM, respectively, the OSNRs required for systems using differential code were found to be 18.1, 20.4, and 22.8 dB. The essential OSNRs for networks with gray-coded are found to be 14.1, 17.2, and 18.8 dB for PM-QPSK, PM-8-QAM, and PM-16-QAM, respectively. This leads to greater OSNR needs for modulation formats with larger bits per symbol. In comparison to differential-coded systems, gray-coded modulations were shown to require less OSNR to be maintained.

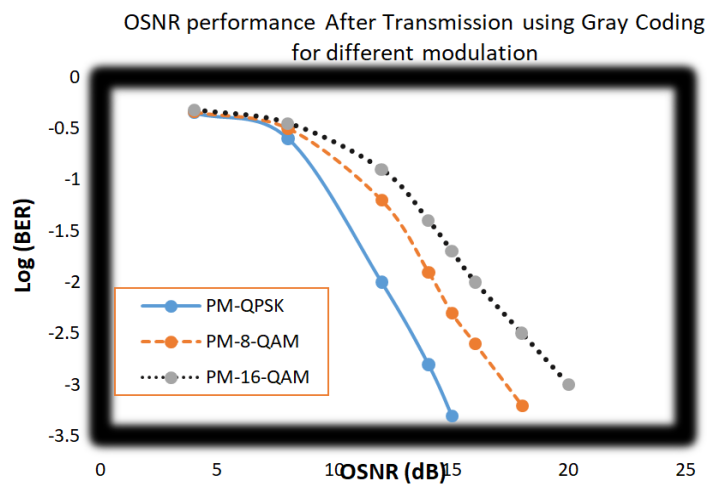

Figure 11 OSNR performance of transmitted signal using Gray Coding

### 6.1. Fixed Grid Network System Capacity And Spectral Efficiency

Figure 2 shows how a fixed grid network uses the WDM system and it may boost system capacity by concurrently transmitting numerous bit streams over the same cable. when an L-length fiber is used to simultaneously broadcast N channels with bit rates *B1, B2*,..., and *B$_N$*. Equation 2[19] expresses the WDM link's overall bit rate. For equal bit rates, the system capacity is increased by a factor of N.

$$B_T = B_1 + B_2 + \cdots\cdots\cdots + B_N \qquad \textit{(Eq.2)}$$

The most crucial design factors for a WDM system are the number of channels N, the bit rate B at which each channel operates, and the frequency separation $\Delta v_{ch}$ between adjacent channels. System capacity is denoted by the term NB. The total bandwidth consumed by a fixed grid network system is denoted by the product $N \times \Delta v_{ch}$. For WDM systems, the standard way to introduce the idea of spectral efficiency is expressed in equation 3[19].

$$\eta_s = B/\Delta v_{ch} \qquad \textit{(Eq.3)}$$

Where Bitrate (B) for PM-QPSK is 100 Gbps and frequency separation $\Delta v_{ch}$ is 0.4 nm which is 50 GHz, so spectral efficiency is 2 (bits/s)/Hz. Similarly, for PM-8- QAM, and PM-16 QAM is 3 (bits/s)/Hz and 4 (bits/s)/Hz respectively. Fig. 13 shows the spectral efficiencies for different data rates by using effective modulation schemes.

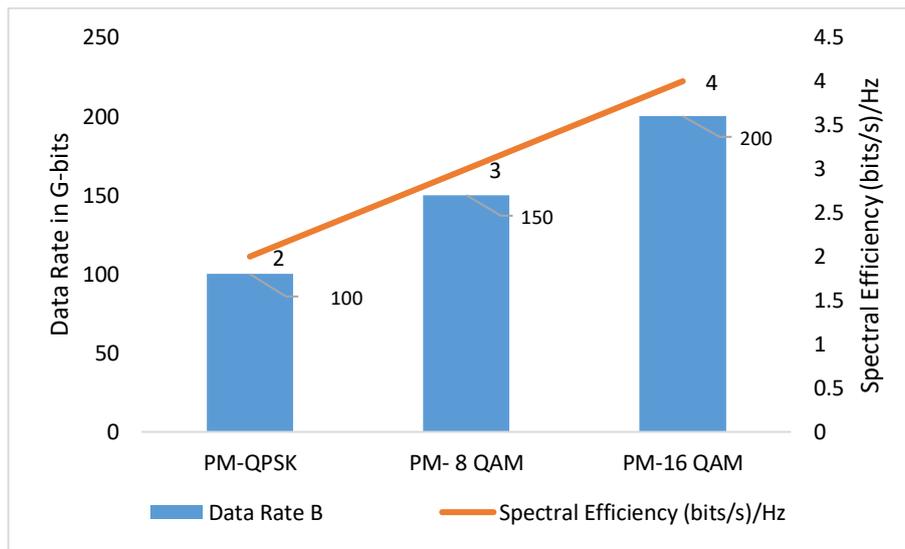

Figure 12. Spectral efficiencies vs data rate for the different modulation schemes

It indicates that an effective modulation scheme gives better spectral efficiency. Figure 14 illustrates the system capacity and optical reach for the different modulation schemes as counterpart optical reach will decrease for the effective modulation scheme but system capacity will be increasing drastically. DWDM system is used for a fixed grid network so that number of channels (N) is 80-90 and an equal bitrate (B) is used for each channel which is 100 Gbps so the system capacity of the network will be 8 -9 Tbps for PM -QPSK.  Similarly, system capacity for PM -8QAM, and PM -16QAM will be 12-13.5 Tbps and 16-18 Tbps for a data rate of 150 Gbps and 200 Gbps respectively.

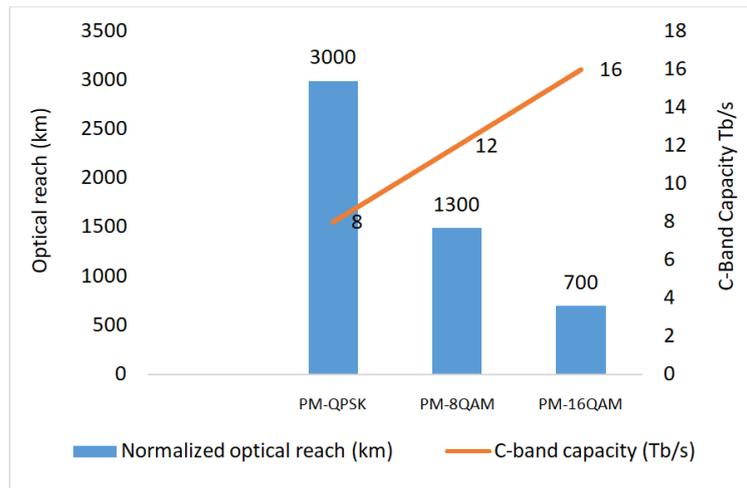

Figure 13 System capacity Vs Optical reach for the different Modulation schemes

The gray-coded IQM-based optical modulation formats are performance compared in Table 5 using three research studies. It has been noted that employing single carrier transmission, the proposed designs exhibit a greater figure of merit, it primarily discusses bit rate and length. Although PDM-16-QAM has a higher data rate than the proposed architecture, the required OSNR is much higher.

Table 5 Performance evaluation of various modulation types with reference to relevant literature

| Parameters | [49] | [50] | [51] | Proposed Work |
|---|---|---|---|---|
| Multiplexed, Modulation and Techniques used | PDM, QPSK, 16-QAM, and DSP | PDM, 16-QAM, WDM, and DSP | DP-16-QAM, PCS and DSP | IQM, PDM, WDM, PM-QPSK, PM-8-QAM, PM-16-QAM, and DSP |
| Number of channels | 1 | 5 | 2 | 80 - 90 |
| Spectral efficiency b/s/Hz | NA | 7.6 | 4 | 2 (PM-QPSK) 3 (PM-8-QAM), 4 (PM-16-QAM) |
| The maximum data rate, Gbps | 112 | 5X256 | 2X200 | 100 (PM-QPSK), 150 (PM-8-QAM), 200 (PM-16-QAM) |
| Maximum SMF length, km | 80-960 | 80 | 90-727 | 3000 (PM-QPSK), 1300 (PM-8-QAM), 700 (PM-16-QAM) |
| Required OSNR, dB | 12.8 (QPSK), 17.1 (16-QAM) | 36 | NA | 14.1 (PM-QPSK), 17.5 (PM-8-QAM), 19.2 (PM-16-QAM) |
| BER | $3.8 \times 10^{-3}$ | $4.5 \times 10^{-3}$ | $4.93 \times 10^{-3}$ | $2 \times 10^{-3}$ |
| Each Channel spacing (GHz) | 50 | 29 | 50 | 50 |
| The total capacity of C-Band(~1530–1565 nm) in Tbps | NA | NA | NA | 8 - 9 (PM-QPSK), 12 -13.5 (PM-8-QAM), 16 - 18 (PM-16-QAM) |

NA: Not available;    PCS: Probabilistic Constellation Shaping

The proposed work mainly includes IQM, PDM, and WDM with different modulation techniques.

It also uses gray coding while encoding the data, which helps the receiver section to retrieve the data. These enhanced performances are the consequence of gray coding, optimized component settings, EDFA, sophisticated DSP compensation algorithms, and straightforward IQ-based higher-order optical modulations. C-band total capacity is studied, to understand the upcoming capacity demand.

# 7. CONCLUSIONS

C-Band spectral and system capacity can be improved by using sophisticated optical fiber but in this work, we kept under-laying cable as it is to nullify the cost consumption. We have improved the system capacity by using different information coding. Especially differential and gray coding is used to overcome bits errors in the receiver and improve the ONSR performance.Different modulation techniques are tested under the fixed grid network and calculated its system capacity and spectral efficiency by adopting both codes. It shows the spectral efficiency increases as we go for higher modulation but transmission distance reduces. PM-QPSK is used for long-distance communication and PM-QAM for short distances. Depending upon the application an appropriate modulation needs to be chosen for better performance under a fixed grid network. To account for fiber attenuation, EDFA with an 8 dB gain is utilized at regular intervals of 40 km SMF. The receiver's DSP unit is used to adjust for a variety of signal faults caused by fiber impairments, including CD, PMD, and Kerr non-linearity. This allows for maximum transmission distances of 3000, 1300, and 700 km for PM-QPSK, PM-8-QAM, and PM-16-QAM, respectively, while maintaining an acceptable BER. The fixed grid network's system capacity will be 8–9 Tbps, 12–13.5 Tbps, and 16–18 Tbps for data rates of 100Gbps, 150Gbps, and 200Gbps, respectively. However, it's important to note that differential encoding can be sensitive to errors, especially if there are significant variations between consecutive values. In cases where the changes are abrupt or large, the cumulative effect of differential encoding can lead to distortion or inaccuracies in the decoded data.Overall, Gray code is a good encoding method that aids in addressing binary value transition concerns, making it helpful in circumstances where precision, mistake detection, or smooth transitions are crucial.Grey-coded systems performed better than differential-coded systems in terms of OSNR requirements for the investigated modulation schemes when the same data rates, transmission distances, and fiber launch optical powers were taken into account. Grey-coded systems provide an OSNR improvement of 4, 3.2, and 4dB for PM-QPSK, PM-8-QAM, and PM-16-QAM, respectively, after SMF transmission. The proposed system is best to transmit 100 Gbps data rate with 50 GHz channel spacing.

We noticed that for lesser data rate (10 Gbps) requests same 50 GHz channel spacing is used so it's not an effective way to utilize the channel spacing. A standard transmission data rate of more than 100Gbps is currently being considered by the datacom and telecom industries, and 400Gbps is receiving a lot of attention. The spectral width occupied by 400 Gbps in standard modulation formats is too wide to fit into the 50 GHz ITU grid. To overcome such issues new techniques need to be adopted. The solution is the Elastic optical network (EON)which mainly deals with the allocation of the channels and handling the channel spacing. EON performance is not studied under practical consideration. In the future, one can develop the Flexi-grid network to achieve effectively utilize channel spacing and to understand its system capacity.

# CONFLICTS OF INTEREST

The authors declare no conflict of interest.

# REFERENCES

[1] E. Agrell *et al.*, "Roadmap of optical communications," *J. Opt. (United Kingdom)*, vol. 18, no. 6, 2016.

[2] C. G. Bell, A. N. Habermann, J. McCredie, R. Rutledge, and W. Wulf, *Computer Networkking: A Top-Down Approach*, vol. 3, no. 5. 1970.

[3] M. Chen, S. Mao, and Y. Liu, "Big data: A survey," *Mob. Networks Appl.*, vol. 19, no. 2, pp. 171–209, 2014.

[4] T. Horvath, P. Munster, V. Oujezsky, and N. H. Bao, "Passive optical networks progress: A tutorial," *Electron.*, vol. 9, no. 7, pp. 1–31, 2020.

[5] "ITU-T Recommendation G.694.2:Spectral grids for WDM applications: CWDM wavelength grid," *Int. Telecomunication Union*, p. 12, 2003.

[6] K. N. S. Rajiv Ramaswami, *Optical networks a practical perspective*, no. 2. 1998.

[7] F. Q. Kareem *et al.*, "A Survey of Optical Fiber Communications: Challenges and Processing Time Influences," *Asian J. Res. Comput. Sci.*, vol. 7, no. 4, pp. 48–58, 2021.

[8] J. Thangaraj, "Review and analysis of elastic optical network and sliceable bandwidth variable transponder architecture," vol. 57, no. 11, 2019.

[9] S. P. Singh, S. Sengar, R. Bajpai, and S. Iyer, "Next-generation variable-line-rate optical WDM networks: Issues and challenges," *J. Opt. Commun.*, vol. 34, no. 4, pp. 331–350, 2013.

[10] "ITU-T Recommendation G.694.1. Spectral grids for WDM applications: DWDM frequency grid," *Int. Telecomunication Union*, 2012.

[11] T. Mizuno and Y. Miyamoto, "High-capacity dense space division multiplexing transmission," *Opt. Fiber Technol.*, 2016.

[12] Cisco, "Cisco Annual Internet Report - Cisco Annual Internet Report (2018 - 2023) White Paper," pp. 1–41, 2020.

[13] D. Kakati and S. C. Arya, "Performance of grey-coded IQM-based optical modulation formats on high-speed long-haul optical communication link," *IET Commun.*, vol. 13, no. 18, pp. 2904–2912, 2019.

[14] R. A. Pagare, S. Kumar, and A. Mishra, "Design and analysis of hybrid optical distribution network for worst-case scenario of E2-class symmetric coexistence 80 Gbps TWDM NG-PON2 architecture for FTTX access networks," *Optik (Stuttg).*, vol. 228, p. 166168, 2021.

[15] V. N. Index, C. Vni, and C. Vni, "Cisco Visual Networking Index : Forecast and Trends: 2017-2022," pp. 1–38, 2019.

[16] R. A. Pagare, A. Mishra, and S. Kumar, "Analytical modeling and impact analysis on multichannel symmetric optical and wireless NG-PON2 networks of CD, SPM, XPM and FWM impairments," *Optik (Stuttg).*, vol. 252,168573, 2022.

[17] Cisco forecast, "Cisco visual networking index (VNI) global mobile data traffic forecast update, 2017-2022 white paper," *Comput. Fraud Secur.*, pp. 3–5, 2019.

[18] R. A. Pagare, A. Mishra, and S. Kumar, "Impairment strained analytical modeling evaluation and cross-talk analysis of symmetric and coexistent channels for extended class-1 NG-PON2 access network," *Opt. Quantum Electron.*, vol. 54, no. 11, pp. 0–21, 2022.

[19] F. Edition and G. P. Agrawal, *Fiber-Optic Communiction System*, Fourth. A JOHN WILEY & SONS, INC., PUBLICATION, 2010.

[20] U. B. Pre- *et al.*, "64-Tb/s, 8 b/s/Hz, PDM-36QAM Transmission Over 320 km Using Both Pre- and Post-Transmission Digital Signal Processing," *J. Light. Technol.*, vol. 29, no. 4, pp. 571–577, 2011.

[21] F. Li, X. Xiao, J. Yu, J. Zhang, and X. Li, "Real-time direct-detection of quad-carrier 200Gbps 16QAM-DMT with directly modulated laser," *Eur. Conf. Opt. Commun. ECOC*, vol. 2015-Novem, no. 1, pp. 16–18, 2015.

[22] K. Kikuchi, "Coherent optical communication technology," in *016 21st OptoElectronics and Communications Conference, OECC 2016 - Held Jointly with 2016 International Conference on Photonics in Switching, PS 2016*, 2016.

[23] S. A. Li *et al.*, "Enabling Technology in High-Baud-Rate Coherent Optical Communication Systems," *IEEE Access*, vol. 8, pp. 111318–111329, 2020.

[24] C. Xia, "Optical Fibers for Space-Division Multiplexed Transmission and Networking," *Phd*, 2015.

[25] C. Wang *et al.*, "Integrated lithium niobate electro-optic modulators operating at CMOS-compatible voltages," *Nature*, vol. 562, no. 7725, pp. 101–104, 2018.

[26] D. Kakati and S. C. Arya, "A full-duplex optical fiber/wireless coherent communication system with

digital signal processing at the receiver," *Optik (Stuttg.)*, vol. 171, no. May, pp. 190–199, 2018.

[27] M. Adarsha and S. Malathi, "Effective utilization of channel spacing in Elastic Optical Network for 400 Gb/s transmissions," in *Futuristic Communication and Network Technologies. VICFCNT 2020. Lecture Notes in Electrical Engineering, vol 792.*, 2022, 1st ed., pp. 619–626.

[28] L. Yue, D. Kong, Y. Li, J. Pang, M. Yu, and J. Wu, "A novel demultiplexing scheme for Nyquist OTDM signal using a single IQ modulator," in *2016 21st OptoElectronics and Communications Conference, OECC 2016 - Held Jointly with 2016 International Conference on Photonics in Switching, PS 2016*, 2016, vol. 1, pp. 26–28.

[29] L. Li, G. Xiao-Bo, and L. Jing, "Analysis of Performance for 100 Gbit/s Dual-Polarization QPSK Modulation Format System," *J. Opt. Commun.*, vol. 37, no. 1, pp. 93–101, 2016.

[30] J. A. Bebawi, I. Kandas, M. A. El-Osairy, and M. H. Aly, "A comprehensive study on EDFA characteristics: Temperature impact," *Appl. Sci.*, vol. 8, no. 9, 2018.

[31] B. Batagelj, V. Janyani, and S. Tomažič, "Research challenges in optical communications towards 2020 and beyond," *Inf. MIDEM*, vol. 44, no. 3, pp. 177–184, 2014.

[32] V. A. Thomas, M. El-Hajjar, and L. Hanzo, "Performance improvement and cost reduction techniques for radio over fiber communications," *IEEE Commun. Surv. Tutorials*, vol. 17, no. 2, pp. 627–670, 2015.

[33] Z. Wang, A. Yang, P. Guo, and P. He, "OSNR and nonlinear noise power estimation for optical fiber communication systems using LSTM based deep learning technique," *Opt. Express*, vol. 26, no. 16, p. 21346, 2018.

[34] D. Kedar and S. Arnon, "Urban optical wireless communication networks: The main challenges and possible solutions," *IEEE Commun. Mag.*, vol. 42, no. 5, pp. 20–22, 2004.

[35] D. Lavery, D. Ives, G. Liga, A. Alvarado, S. J. Savory, and P. Bayvel, "The Benefit of Split Nonlinearity Compensation for Single-Channel Optical Fiber Communications," *IEEE Photonics Technol. Lett.*, vol. 28, no. 17, pp. 1803–1806, 2016.

[36] J. Pfeifle *et al.*, "Optimally coherent Kerr combs generated with crystalline whispering gallery mode resonators for ultrahigh capacity fiber communications," *Phys. Rev. Lett.*, vol. 114, no. 9, pp. 1–5, 2015.

[37] C. F. Lam, H. Liu, B. Koley, X. Zhao, V. Kamalov, and V. Gill, "Fiber Optic Communication Technologies: What's Needed for Datacenter Network Operations," *IEEE Communications Magazine*, no. July, pp. 32–39, 2010.

[38] C. Kachris and I. Tomkos, "A survey on optical interconnects for data centers," *IEEE Commun. Surv. Tutorials*, vol. 14, no. 4, pp. 1021–1036, 2012.

[39] E. Temprana *et al.*, "Overcoming Kerr-induced capacity limit in optical fiber transmission," *Science (80-. ).*, vol. 348, no. 6242, pp. 1445–1448, 2015.

[40] J. Zhang *et al.*, "Functional-link neural network for nonlinear equalizer in coherent optical fiber communications," *IEEE Access*, vol. 7, pp. 149900–149907, 2019.

[41] B. Ragchaa and K. Kinoshita, "Spectrum Sharing between Cellular and Wi-Fi Networks Based on Deep Reinforcement Learning," *Int. J. Comput. Networks Commun.*, vol. 15, no. 1, pp. 123–143, 2023.

[42] ITU, "ITU G.694.1 (06/2002): Spectral grids for WDM applications: DWDM frequency grid," 2002.

[43] D. Sharma, S. Bajpai, and Y. K. Prajapati, "Next generation PON using PM-BPSK and PM-QPSK modulation," *IMPACT 2017 - Int. Conf. Multimedia, Signal Process. Commun. Technol.*, pp. 10–12, 2018.

[44] M. S. Faruk and S. J. Savory, "Digital Signal Processing for Coherent Transceivers Employing Multilevel Formats," *J. Light. Technol.*, vol. 35, no. 5, pp. 1125–1141, 2017.

[45] M. Stepanovsky, "A Comparative Review of MEMS-based Optical Cross-Connects for All-Optical Networks from the Past to the Present Day," *IEEE Commun. Surv. Tutorials*, vol. PP, no. XXX, p. 1, 2019.

[46] M. R. Sena, P. J. Freire, L. D. Coelho, A. F. Santos, A. Napoli, and R. C. Almeida, "Novel evolutionary planning technique for flexible-grid transmission in optical networks," *Opt. Switch. Netw.*, vol. 43, no. February 2021, p. 100648, 2022.

[47] L. S. de Sousa and A. C. Drummond, "Metropolitan Optical Networks: A Survey on New Architectures and Future Trends," pp. 1–59, 2022.

[48] H. Huang, "Hybrid Flow Data Center Network Architecture Design and Analysis," Massachusetts Institute of Technology, 2017.

[49] M. Mehra, H. Sadawarti, and M. L. Singh, "Performance analysis of coherent optical communication system for higher order dual polarization modulation formats," *Optik (Stuttg.)*, vol. 135, pp. 174–179, 2017.

[50] Y. Zhu, M. Jiang, Z. Chen, and F. Zhang, "Terabit Faster-Than-Nyquist PDM 16-QAM WDM Transmission with a Net Spectral Efficiency of 7.96 b/s/Hz," *J. Light. Technol.*, vol. 36, no. 14, pp. 2912–2919, 2018.

[51] Y. R. Zhou and K. Smith, "Practical Innovations Enabling Scalable Optical Transmission Networks: Real-World Trials and Experiences of Advanced Technologies in Field Deployed Optical Networks," *J. Light. Technol.*, vol. 38, no. 12, pp. 3106–3113, 2020.

## AUTHORS

**Mr. Adarsha M,** he obtained Diploma in Electronics and Communication (E&C) Engineering (2009) from N.R.A.M Aided Polytechnic Nitte. He received B.E and M.Tech degrees in Engineering from the Visvesvaraya Technological University (VTU) in 2012 and 2014, respectively. He is currently pursuing his Ph.D. in Optical Communication in the department of E&C Engineering, M. S. Ramaiah University of Applied Sciences, (MSRUAS) Bangalore, India. He worked as a Temporary Assistance Professor in the department of E&C, at the National Institute of Technology Karnataka (NITK) surthkal. He completed several online courses related to optical networks namely an introduction to optical networks, and Dense Wavelength Division Multiplexing (DWDM) Networks. He also did a Diploma in fiber optic communication technology. He got the best research paper award at the International Conference on Futuristic Communication and Network Technologies (2020). He has participated in the Design Competition on "Optical Communication & Photonic Design" using Optiwave Photonic Design & Simulation Tools, organized by the IEEE Photonics Students Chapter, and secured First Prize (2021).
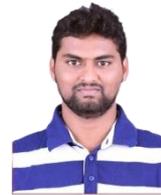

**Dr. Malathi S** is a Professor and Head of the Department of Electronics and Communication Engineering, M. S. Ramaiah University of Applied Sciences, (MSRUAS) Bangalore, India. She obtained her B.E in Electronics and Communication Engineering (1996), M.E in Microwave and Optical Engineering (1999), and Ph.D. (2012) from the Indian Institute of Science (IISc) Bangalore.She is a thesis Advisor for undergraduate, postgraduate (MS), and Ph.D. students. She has been the Chairperson of the IISc — IEEE student branch in 2008, an Executive Committee member, of the IEEE Photonics Society (2013 – 2018), and its Chair-Elect (2020) for the Bangalore section. Awarded a student scholarship by SPIE — International Society for Optics and Photonics US (2008), she also received the Professor Selvarajan grant for international travel from the Indian Institute of Science in the same year. She has published and co-authored several papers in International Journals as well as Conferences on subjects relating to Optical networks, and optic resonators for bio-sensing applications. Dr. Malathi is a member of SPIE and a Senior Member of IEEE Photonics and a Life member of IETE.
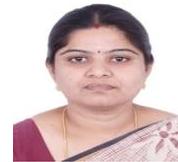

**Dr. Santosh Kumar** received a Ph.D. degree from the Indian Institute of Technology (Indian School of Mines) Dhanbad, Dhanbad, India in 2014. He is currently an Associate Professor at the School of Physics Science and Information Technology, Liaocheng University, Liaocheng, China. He has published more than 230 high-quality research articles in national and international SCI journals and conferences. He has presented numerous articles at conferences held in China, India, Belgium, and the USA. He has recently published two scholarly books 2D Materials for Surface Plasmon Resonance-Based Sensors (CRC Press, 2021) and Optical Fiber-Based Plasmonic Biosensors: Trends, Techniques, and Applications (CRC Press, 2022). He has also authored a textbook Fiber Optic Communication: Optical Waveguides, Devices, and Applications (University Press, 2017). He has reviewed over 1300 SCI journals published by IEEE, Elsevier, Springer, OPTICA, SPIE, Wiley, ACS, and Nature up to this point. Dr. Kumar is a Fellow of SPIE and a Senior Member of IEEE and OPTICA. He is an OPTICA Traveling Lecturer. He has received the "2022 Best Performing Associate Editor" Award from IEEE SENSORS JOURNAL. He is also the Chair of the Optica Optical Biosensors Technical Group. He has given numerous invited speeches and serves as the session chair for IEEE conferences. He has served as an Associate Editor for IEEE Sensors Journal, IEEE Access, IEEE Transactions on Nanobioscience, Frontiers of Physics, and Biomedical Optics Express.
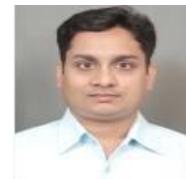